\documentstyle[amssymb,aps,12pt]{revtex}

\begin{document}
\draft
\author{O. B. Zaslavskii}
\address{Department of Mechanics and Mathematics, Kharkov V.N. Karazin's National\\
University, Svoboda\\
Sq.4, Kharkov 61077, Ukraine\\
E-mail: aptm@kharkov.ua}
\title{Thermodynamics of black holes with an infinite effective area of a horizon }
\maketitle

\begin{abstract}
In some kinds of classical dilaton theory there exist black holes with (i)
infinite horizon area $A$ or infinite $F$ (the coefficient at curvature in
Lagrangian) and (ii) zero Hawking temperature $T_{H}$. For a generic static
black hole, without an assumption about spherical symmetry, we show that
infinite $A$ is compatible with a regularity of geometry in the case $%
T_{H}=0 $ only. We also point out that infinite $T_{H}$ is incompatible with
the regularity of a horizon of a generic static black hole, both for finite
or infinite $A$. Direct application of the standard Euclidean approach in
the case of an infinite ''effective'' area of the horizon $A_{eff}=AF$ leads
to inconsistencies in the variational principle and gives for a black hole
entropy $S$ an indefinite expression, formally proportional to $T_{H}A_{eff}$%
. We show that treating a horizon as an additional boundary (that is, adding
to the action some terms calculated on the horizon) may restore
self-consistency of the variational procedure, if $F$ near the horizon grows
not too rapidly. We apply this approach to Brans-Dicke black holes and
obtain the same answer $S=0$ as for ''usual'' (for example,
Reissner-Nordstr\"{o}m) extreme classical black holes. We also consider the
exact solution for a conformal coupling, when $A$ is finite but $F$ diverges
and find that in the latter case both the standard and modified approach
give rise to an infinite action. Thus, this solution represents a rare
exception of a black hole without nontrivial thermal properties.
\end{abstract}

\pacs{PACS numbers: 04.70.Dy, 04.50+h}


\section{Introduction}

At present, black hole thermodynamics is put on a firm basis due to the
elaborated Euclidean action formalism. However, some gaps in important
issues still persist here. In particular, it concerns gravity with dilaton
(scalar) field. It was shown that some kinds of such a theory (in
particular, Brans-Dicke theory \cite{jor}) predict quite unusual objects,
having no analogs in general relativity - black holes with an infinite
surface area of an event horizon - \cite{krori} - \cite{cold}. The natural
question about thermodynamical interpretation (first of all, entropy) arises
and an unusual character of such black holes makes it challenging.

Naive application of the Bekenstein - Hawking formula for the black hole
entropy, proportional to the horizon area, would give here a physically
meaningless answer. On the other hand, it turns out that the surface gravity 
$\kappa $ for such black holes and the corresponding Hawking temperature $%
T_{H}=\frac{\kappa }{2\pi }$ are exactly zero. This seemed to suggest
another way of treating black hole thermodynamics, in the spirit of the
already elaborated approach to extreme black holes \cite{ross} - \cite{gib}.
According to it, an arbitrary finite temperature and zero entropy should be
ascribed to classical extreme black holes. As the horizon area does not
enter explicitly here, it would seem, on the first glance, that an unusual
character of black holes with infinite horizon area does not change the
approach. Nevertheless, thorough examination, suggested below, shows that
this is not the case. First, naive calculation of the Euclidean action
leads, as we will see, to the model-dependent contribution for a candidate
on the entropy, that contradicts our expectations, based on knowledge of
universality of black hole physics. Moreover, one cannot guarantee its
finiteness and even non-negativity. Second, it turns out that the standard
expressions for the Euclidean action has the following undesirable feature:
it is not invariant in the case under discussion with respect to conformal
transformation because of some additional horizon contribution, so its form
depends on the conformal frame. Third, if we trace back the variational
principle, from which the solutions under discussion are obtained, it turns
out that the variation of the action contains the terms (including the
normal derivative of the metric) that vanished on the horizon for ''usual''
black holes but persist (and even may become infinite) for solutions
admitting an infinite horizon area.

To resolve these difficulties, we suggest modification of the Euclidean
action that resolves all three problems, inherent to the infinite area case,
at once. In so doing, the third point is central since it concerns the
foundation of the variation principle from which other properties stem. The
modification consists in treating a horizon as an additional boundary that
formally means adding to the action some terms calculated on the horizon.
For ''usual'' extreme black holes these terms are automatically zero, so
this modification agrees with previously obtained results \cite{ross} - \cite
{gib}. Our analysis is valid for any static black holes and is not
restricted to Brans-Dicke black holes, including them only as an example. We
also examine the thermodynamic interpretation of black holes when the
horizon area is finite, but the coupling between dilaton and curvature
diverges.

It is worth noting that an attempt to examine thermodynamic interpretation
of black holes in Brans-Dicke theory was made in \cite{kim}. In our view,
the conclusion, made there, about complete failure of thermodynamic approach
to black holes in Brans-Dicke theory, is incorrect. The range of parameters,
considered in \cite{kim}, corresponds to singular horizons, when it is
obvious in advance that thermodynamics has no physical meaning. However,
there exists another range (omitted in \cite{kim}), where the geometry on
the horizon is regular and thermodynamics of Brans-Dicke black holes is
well-defined (see details below).

\section{Regular horizons with an infinite area}

Before addressing thermodynamics issues, let us proof the following lemma.

{\it Static} {\it black holes with a regular horizon, having and infinite
surface area }$A${\it \ can be extreme only (the surface gravity }$\kappa =0$%
{\it ).}

Consider an arbitrary static spacetime with an event horizon. It is
convenient to use the coordinate system, exploited in \cite{israel}: 
\begin{equation}
ds^{2}=-V^{2}dt^{2}+\rho ^{2}dV^{2}+\gamma _{ab}d\theta ^{a}d\theta ^{b}%
\text{, }a,b=1,2\text{.}  \label{mi}
\end{equation}
Then, it follows from the corresponding geometrical formulas that the
Kretscmann invariant 
\begin{equation}
I\equiv \frac{1}{4}R_{\alpha \beta \gamma \delta }R^{\alpha \beta \gamma
\delta }=G_{i}^{j(3)}G_{j}^{i(3)}+V^{-2}Y_{ij}Y^{ij}\text{, }Y_{ij}\equiv
V_{;i;j}  \label{inv}
\end{equation}
where $^{(3)}$refers to the three-geometry of slices $t=const$, the
covariant derivatives are also taken with respect to three-geometry. Simple
calculations give us 
\begin{equation}
Y_{ij}Y^{ij}=\rho ^{-2}[K_{ab}K^{ab}+2\rho ^{-2}\rho _{,a}\rho ^{,a}+\rho
^{-4}\left( \frac{\partial \rho }{\partial V}\right) ^{2}]\text{,}
\label{yinv}
\end{equation}
where $K_{ab}$ is two-dimensional extrinsic curvature tensor, 
\begin{equation}
K_{ab}=\frac{1}{2\rho }\frac{\partial \gamma _{ab}}{\partial V}\text{.}
\label{yk}
\end{equation}
All indices in the right hand side of (\ref{inv}) and (\ref{yinv}) are
raised and lowed with respect to the three-dimensional metric of spacelike
slices $t=const$ which is positive-definite$.$ Therefore, the first and the
second terms in the right hand side of (\ref{inv}) and (\ref{yinv}) are
non-negative and, therefore, should be finite separately. In particular, it
is true for terms with $K_{ab}K^{ab}$. One can, for example, without the
loss of generality, diagonalize $K_{ab}$ and conclude that both $K_{1}^{1}$, 
$K_{2}^{2}$ as well as the extrinsic curvature $K^{(2)}=K_{1}^{1}+K_{2}^{2}$
should be finite. In our context, below we will take advantage of the
finiteness of the second term in (\ref{inv})$.$

Now observe that it follows from (\ref{yk}) that 
\begin{equation}
\frac{\partial \eta }{\partial V}=\langle K^{(2)}\rho \rangle \text{, }\eta
\equiv \ln A\text{, }\langle ...\rangle =A^{-1}\int d\theta ^{1}d\theta ^{2}%
\sqrt{\gamma }(...)\text{, }\gamma =\det (\gamma _{ab})\text{,}  \label{kr}
\end{equation}
where $A=\int d\theta ^{1}d\theta ^{2}\sqrt{\gamma }$ is the surface area of
the slice with a constant $V$. Actually, the quantity $V$ enumerates
equipotential surfaces $-g_{00}=V^{2}=V_{0}^{2}=const$. If we take some
surface $f(V$, $y^{1}$, $y^{2})=0$, find the components $n_{\mu }\backsim
\partial _{\mu }f$ of the vector, orthogonal to this surface and calculate $%
n_{\mu }n^{\mu }$, one can easily see that, for the metric (\ref{mi}) and
the choice $f=V^{2}-V_{0}^{2}$, such a quantity on the surface $V=V_{0}$ is
proportional to $V^{2}$. Thus, when $V\rightarrow 0$, the vector $n_{\mu }$
becomes isotropic. This means that the surface under discussion represents a
Killing horizon which, due to the staticity of the metric, coincides with
the event horizon. Thus, the position of the horizon is determined by the
condition $V=0$ and does not involve other metric functions or coordinates.
As $g_{00}=0$ on the horizon and $g_{00}<0$ everywhere in the static region,
as usual (in particular, $g_{00}\rightarrow -1$ at infinity in
asymptotically flat spacetimes), $g_{00}$ as well as the quantity $V$ cannot
vanish in an intermediate region between a horizon and infinity, so one can
choose $V>0$ there (for example, $V=\sqrt{1-\frac{2m}{r}\text{ }}$in the
Schwarzschild case in curvature coordinates).

As one approaches the horizon, $V\rightarrow 0$ and it follows from the
finiteness of $I$ that $\rho _{,a}\rightarrow 0$. Thus, one obtains the
constancy of the surface gravity $\kappa =\rho _{0}^{-1}$, where $\rho
_{0}=\lim_{V\rightarrow 0}\rho $ (the zero law of black hole thermodynamics 
\cite{bar}).

Let a black hole be non-extreme, $\kappa \neq 0$, $\rho _{0}$ is finite.
Then we see from (\ref{inv}), (\ref{yinv}), (\ref{kr}) that $K^{(2)}\sim
a(\theta ^{1},\theta ^{2})V\rightarrow 0$ near the horizon, where $a(\theta
^{1},\theta ^{2})\,$is bounded on the horizon. Taking into account that, by
assumption, $\rho _{0}$ is finite, we obtain that (i) $\frac{\partial \eta }{%
\partial V}\sim V\rightarrow 0$. If we want to have a black hole with an
infinite horizon area, (ii) $\eta \rightarrow \infty $ near the horizon. It
is obvious that both properties (i) and (ii) are mutually inconsistent that
forbids the existence of regular nonextreme black holes with an infinite
horizon area. This completes the proof.

However, if a black hole is extreme, $\kappa =0$, $\rho \rightarrow \infty $%
, $\frac{\partial \eta }{\partial V}\sim V\langle a\rho \rangle $ and we
have a competition of two factors $V$ and $\rho $. This leaves the
opportunity of infinite $A$ that is indeed realized in some
scalar-gravitation theories, as is clear from corresponding exact solution
in the spherically-symmetrical case (\cite{jor}) - (\cite{cold}).

Now we address briefly one more issue. There was made an observation in \cite
{cold0} for arbitrary spherically-symmetrical configurations that an
infinite $T_{H}$ is inconsistent with a regularity of a horizon. Now we
extend this observation to an arbitrary static (not necessarily
spherically-symmetrical) configuration. Let $\kappa =\infty $. This means
that $\rho \rightarrow 0$ near the horizon. It follows from (\ref{inv}), (%
\ref{yinv}) that (a) $x\equiv \rho ^{-2}\rightarrow \infty $, (b) $\frac{%
\partial x}{\partial V}\rightarrow 0$, when $V\rightarrow 0$. It is clear
that (a) and (b) are mutually inconsistent. It is worth noting that this
conclusion is valid irrespective of whether the horizon area is finite or
infinite.

{\it Thus, an arbitrary static black hole with a regular horizon cannot have
an infinite surface gravity (infinite Hawking temperature).}

\section{Standard Euclidean action approach to dilaton black holes}

In this section we rederive basic formulas of black hole thermodynamics for
the presence of a scalar field (dilaton). We use the generalization of the
Hilbert action to the dilaton case. If the Hamiltonian constraint is taken
into account, the action, as we will see, takes the thermodynamic form with
the local temperature (or its inverse $\beta $)\ on the boundary as a
relevant thermodynamic parameter. In other words, fixing $\beta $ on the
boundary, we work with the canonical ensemble throughout the paper. By
itself, such an extension to dilaton theories is quite direct. Meanwhile, we
will need them for our purposes in general setting. In so doing, I follow
almost the same line as in \cite{can}, where the general canonical approach
to self-gravitating systems was suggested (having generalized previous
observations for the spherically-symmetrical case \cite{york86}).

We restrict ourselves to static spacetimes only. Then the Euclidean metric
can be written as 
\begin{equation}
ds^{2}=b^{2}d\tau ^{2}+g_{ij}dx^{i}dx^{j}\text{,}  \label{m}
\end{equation}
where all functions are independent of $\tau $. Consider the system governed
by the Eucliean gravitation-dilaton action $I_{gd}=I_{V}+I_{B}$, where the
bulk part 
\begin{equation}
I_{V}=-\frac{1}{16\pi }\int_{M}d^{4}x\sqrt{-g}[RF(\phi )+V(\phi )(\nabla
\phi )^{2}+U(\phi )]  \label{ac}
\end{equation}
is taken over the manifold $M$, and the term over the boundary $\partial M$%
\begin{equation}
I_{B}=\frac{1}{8\pi }\int_{\partial M}d^{3}x\sqrt{g_{3}}FK  \label{b}
\end{equation}
is necessary to make the variational procedure self-consistent, $K$ is the
second fundamental form of the boundary. If $F=1$, we return to the case of
general relativity. In the Euclidean action integration is performed over
Euclidean time $\tau $, $0\leq \tau \leq \beta _{0}=T_{0}^{-1}$. In what
follows it is assumed that the Euclidean manifold is regular and does not
contain conical singularities. This implies that we consider either
non-extreme black holes with $T_{0}=T_{H}\,$or extreme ones ($T_{H}=0$).

One can derive from (\ref{ac}) the field equations 
\begin{equation}
Z_{\mu }^{\nu }\equiv 2FG_{\mu }^{\nu }+2(\delta _{\mu }^{\nu }\square
F-\nabla _{\mu }\nabla ^{\nu }F)-U\delta _{\mu }^{\nu }+2V\nabla _{\mu }\phi
\nabla ^{\nu }\phi -\delta _{\mu }^{\nu }V(\nabla \phi )^{2}=0\text{,}
\label{fe}
\end{equation}
$\nabla _{\mu }$ is the operator of four-dimensional covariant
differentiation with respect to the metric $g_{\mu \nu }$.

Take into account that 
\begin{equation}
R=-2G_{\tau }^{\tau }-2\frac{\Delta _{3}b}{b}\text{,}  \label{r}
\end{equation}
where Laplacian should be calculated with respect to the three-dimensional
metric $g_{ik}$. The extrinsic curvature $K=-\nabla _{\mu }n^{\mu }$, where $%
n^{\mu }$ is vector, orthogonal to the boundary surface and pointed outward.
Then, writing $K=-\frac{(b\sqrt{g_{3}}n^{\mu })_{,\mu }}{b\sqrt{g_{3}}}$, we
easily obtain 
\begin{equation}
K=K^{(2)}-\frac{b_{,i}n^{i}}{b}\text{,}  \label{k2k}
\end{equation}
comma denotes ordinary derivative. Here $K^{(2)}$ is the extrinsic
curvature, measured with respect to a three-dimensional metric $g_{ij}$, $%
K^{(2)}=-\frac{\left( n^{i}\sqrt{g_{3}}\right) _{,i}}{\sqrt{g_{3}}}$ .
Applying the Gauss theorem, after some algebra we obtain 
\begin{equation}
I_{gd}=\frac{\beta _{0}}{16\pi }\int d^{3}x\sqrt{g_{3}}bZ_{0}^{0}+\int_{B}d%
\sigma \beta \varepsilon +Y_{1}-Y_{2}\text{,}  \label{itot}
\end{equation}
\begin{equation}
\varepsilon =-\frac{\left( n^{i}F\right) _{;i}}{8\pi }=-\frac{\left( n^{i}%
\sqrt{g_{3}}F\right) _{,i}}{\sqrt{g_{3}}}\text{,}  \label{e}
\end{equation}
\begin{equation}
Y_{1}=\frac{\beta _{0}}{8\pi }\int_{H}d\sigma bF_{,i}n^{i}\text{,}
\label{y1}
\end{equation}
\begin{equation}
Y_{2}=\frac{\beta _{0}}{8\pi }\int_{H}d\sigma \frac{\partial b}{\partial l}F%
\text{,}  \label{y2}
\end{equation}
where $d\sigma $ is the element of a two-dimensional surface, $\frac{%
\partial b}{\partial l}=b_{,i}n^{i}$ is the normal derivative, $\beta =\beta
_{0}b$ is the inverse local Tolman temperature, $\varepsilon $ has the
meaning of quasilocal energy density for gravitation-dilaton system and
directly generalizes the corresponding formula for pure gravitation case due
to the factor $F$. The index ''B'' refers to a physical boundary (it is
supposed that we have a black hole enclosed in a cavity), ''H'' is related
to the event horizon.

Now consider separately two cases. First, let on the horizon both $F$ and
the surface area $A$ be finite (''normal'' case). Then $Y_{1}=0$ due to the
factor $b$. On the horizon $\frac{\partial b}{\partial l}\rightarrow \kappa $%
, where $\kappa $ is a surface gravity, constant on it due to the zero law
of black hole thermodynamics. As a result, we get the thermodynamic form for
the action 
\begin{equation}
I_{gd}=\int_{B}d\sigma \beta \varepsilon -S\text{,}  \label{i}
\end{equation}
where the black hole entropy is identified with $Y_{2}$, its value is equal
to 
\begin{equation}
S=\frac{A_{eff}}{4}\frac{T_{H}}{T_{0}}\text{,}  \label{sef}
\end{equation}
$T_{H}\equiv \frac{\kappa }{2\pi }$ is the Hawking temperature, $T_{0}=\beta
_{0}^{-1}$, $A_{eff}=AF_{h}$, $F_{h}$ is the value of $F$ on the horizon.
The formula (\ref{sef}) embraces at once two kinds of regular topologies -
nonextreme black holes with $T_{H}=T_{0}$ and extreme ones.

(i) In the first case we obtain for the black hole entropy the
Bekenstein-Hawking value, generalized to the presence of a scalar field: 
\begin{equation}
S=\frac{A_{eff}}{4}\text{.}  \label{sne}
\end{equation}

(ii) If a black hole is extreme ($T_{H}=0$), and if we identified the
Euclidean time with an arbitrary finite period $T_{0}^{-1}$($T_{0}\neq 0$),
we get from (\ref{sef}) that $S=0\,$in accordance with prescription of \cite
{ross} - \cite{gib}. The basic formulas remain practically intact if the
terms, say, with electromagnetic field are included into the scheme.

However, what we will be interested in is not this slight generalization of
the action formalism to the scalar field, but the special cases of infinite $%
A_{eff}$ which exist due to this field entirely and have no analogs in
general relativity. We will see that direct application of the above
formulas fails and we are led to some modification of the boundary terms in
the action.

\section{Conformal transformations}

Consider the transformations 
\begin{equation}
g_{\mu \nu }=e^{2\psi }\bar{g}_{\mu \nu }\text{, }\sqrt{g}=\sqrt{\bar{g}}%
e^{4\psi }\text{, }\psi =\psi (\phi )\text{.}  \label{con}
\end{equation}
Then 
\begin{equation}
R=e^{-2\psi }\{\bar{R}-6[(\bar{\nabla}\psi )^{2}+\bar{\Box}\psi ]\}.
\end{equation}
Now we should take into account that $n^{\mu }=e^{-\psi }\bar{n}^{\mu }$, 
\begin{equation}
K=-\frac{(\bar{n}^{\mu }\sqrt{\bar{g}}e^{3\psi })_{,\mu }}{\sqrt{\bar{g}}%
e^{4\psi }}=e^{-\psi }\bar{K}-3e^{-\psi }\psi _{,\mu }\bar{n}^{\mu }\text{.}
\end{equation}
We see that $I_{gd}(g_{\mu \nu },F,U,V)=I_{gd}(\bar{g}_{\mu \nu },\bar{F},%
\bar{U},\bar{V})+\Delta I$, where

\begin{equation}
\bar{F}=Fe^{2\psi }\text{, }\bar{U}=Ue^{4\psi }\text{, }\bar{V}=e^{2\psi
}[V+6\psi ^{\prime }(F\psi ^{\prime }+F^{\prime })]\text{,}  \label{f}
\end{equation}
prime denotes here differentiation with respect to $\phi $, 
\begin{equation}
\Delta I=-\frac{3\beta _{0}}{8\pi }\int_{H}d^{2}x\sqrt{\bar{g}_{2}}\bar{F}%
\bar{b}\psi ^{\prime }\phi _{,i}\bar{n}^{i}\text{,}  \label{ex}
\end{equation}
integration being performed over the horizon surface. For the ''normal''
case (with finite $A_{eff}$) this quantity vanishes due to the factor $b$.
Thus, the total Euclidean action in the form (\ref{ac}), (\ref{b}) is
conformally invariant: $I_{gd}(g_{\mu \nu },F,U,V)=I_{gd}(\bar{g}_{\mu \nu },%
\bar{F},\bar{U},\bar{V})$. Apart from this, the entropy itself is also
conformally invariant. For the nonextreme case (when $S=\frac{A_{eff}}{4}$)
it follows from the fact that $A_{eff}$ is invariant due to (\ref{con}), (%
\ref{f}). For the extreme one $S=0$ this invariance becomes obvious. The
conformal invariance of the entropy is physically important because of
intimate links between entropy and conformal structure associated with it
(see for example, the recent work \cite{car} and literature quoted there).

The formula (\ref{i}) can be also rewritten as 
\begin{equation}
I_{eff}=\int_{B}d\sigma _{eff}\beta \varepsilon _{eff}-S\text{,}  \label{i2}
\end{equation}
where $d\sigma _{eff}=d\sigma F$, $S=\frac{A_{eff}}{4}$. 
\begin{equation}
\varepsilon _{eff}=F^{-1}\varepsilon =-\frac{\left( n^{i}F\right) _{;i}}{%
8\pi F}  \label{e2}
\end{equation}

The advantage of the form (\ref{i2}), (\ref{e2}) consists in that it
manifests explicitly invariance of the entropy and action under conformal
transformations (\ref{con}). In so doing, the product $\beta \varepsilon
_{eff}$ by itself is invariant, in contrast to $\beta \varepsilon $.

For the ''anomalous'' case (infinite $A_{eff}$) the situation is more
complicated: we have a play of two factors $b$ and $A_{eff}$ that may result
in unsatisfactory behavior of the action under conformal transformations. It
is worth stressing that we always may perform the conformal transformation
that makes $\bar{F}=1$ or $\bar{F}=-1$ but this transformation leads to the
system, non-equivalent to the original one in the anomalous case since the
new metric may become singular because of infinite or zero $F$ on the
horizon.

\section{Self-consistency of variational procedure and modified action}

The thermodynamic interpretation follows from (\ref{i}) - (\ref{sne}),
provided both $F$ and $A$ are finite. However, our goal is just to handle
the situation when this is not the case. Moreover, ''standard'' pure scalar
(dilaton) black holes are ruled out by general no-hair theorems (see the
review \cite{bek98} and literature quoted there), so in the absence of
electromagnetic (or other gauge fields) what remains is the anomalous case
only. The existence of black holes in the Brans- Dicke theory, without
contradiction to the aforementioned theorems, stems from the fact that for
them $A\rightarrow \infty $ and $F\rightarrow 0$ or $F\rightarrow \infty $,
so the arguments of \cite{bek98} do not apply. As a result, it turns out
that (for example, for the Brans-Dicke theory) we get in (\ref{y1}), (\ref
{y2}) the competition of three factors ($F$, $A$ and $b$ or $b_{,i}$)$.$
Therefore, one can identify in an obvious way the entropy neither with (\ref
{sne}) nor with $S=0$, typical of ''normal'' extreme black holes. For the
same reasons, there is no guarantee that $\Delta I=0$, thus leaving very
undesirable dependence of the form of the action on the conformal frame.

To elucidate the origin of the difficulties, let us return to the action
principle as a starting point and trace back, how the Hamiltonian constraint
appears as a result of variation with respect to $\beta $ (or $b$)$.$
Preliminarily, take into account that $2G_{0}^{0}=-R_{3}$, where $R_{3}$ is
the curvature of the slice $\tau =const$. It is easy to show from (\ref{fe})
that 
\begin{equation}
Z_{0}^{0}=-FR_{3}+2\Delta _{3}F-U-V\left( \nabla \phi \right) ^{2}
\label{z00}
\end{equation}
does not contain $b$. Then, assuming that $\beta _{0}=const$ (for example,
one may choose $\beta _{0}=2\pi $), we get immediately from (\ref{itot})
that 
\begin{equation}
\delta I_{gd}=\frac{\beta _{0}}{16\pi }\int d^{3}x\delta b\sqrt{g_{3}}%
Z_{0}^{0}+\int_{B}d\sigma \delta \beta \varepsilon +\delta Y_{1}-\delta Y_{2}%
\text{.}  \label{di}
\end{equation}
In the normal case $\delta Y_{1}=\frac{\beta _{0}}{8\pi }\int_{H}d\sigma
\delta bF_{,i}n^{i}$ $=0$ since on the horizon $b=0$ is kept fixed, so $%
\delta b=0$ as well, and all other quantities in the integrand remain
finite. Moreover, by a suitable conformal transformation one can achieve $%
F=\pm 1$, so the term $Y_{1}$ does not appear in the action at all. $\delta
Y_{2}=0$ since for extreme topologies $Y_{2}=0$ and for nonextreme ones,
according to (\ref{sef}), (\ref{sne}) $Y_{2}=\frac{A_{eff}}{4}$, the metric
on the horizon being supposed to held fixed. Assuming that $\delta b=0\,$on
the boundary, one gets the Hamiltonian constraint from (\ref{di}).

However, in the anomalous case these arguments do not work for the reasons
explained above. For example, in spite of $\delta b\rightarrow 0$ near the
horizon, the surface element $d\sigma $ or $F_{,i}$ (or both) diverge.
Therefore, to exclude undesirable terms on the horizon with $\delta b$ and $%
\delta \frac{\partial b}{\partial l}$, one is led to kill them by
introducing corresponding counterparts. Namely, let us add on the horizon
the term having the same form as (\ref{b}) (formally, that means that we
treat a horizon as an additional boundary). Then, after some rearrangement,
we get the modified action 
\begin{equation}
\tilde{I}_{gd}=\int_{B}d\sigma \beta \varepsilon -\int_{H}d\sigma \beta
\varepsilon =\int_{B}d\sigma _{eff}\beta \varepsilon _{eff}-\int_{H}d\sigma
_{eff}\beta \varepsilon _{eff}  \label{ith}
\end{equation}

(The minus sign in the second term in (\ref{ith}) arises due to the fact the
outward normal is pointed now into the direction of the horizon.)

We would like to stress that the result $S=0$ is obtained now for the
anomalous case that required automatically some modification of the action
principle. Thus, in combination with the conclusions made in \cite{ross} - 
\cite{gib}, this completes the proof of the property $S=0$ for {\it %
classical }extreme black holes, including here black holes of an infinite
horizon area.

Now, with the modified action, one can easily check that the term (\ref{ex})
does not appear at all and the total modified action, including boundary
terms (recall that now the horizon is an essential piece of boundary) obeys
in the anomalous case the relation $\tilde{I}_{gd}(g_{\mu \nu },F,U,V)=%
\tilde{I}_{gd}(\bar{g}_{\mu \nu },\bar{F},\bar{U},\bar{V})$ similar to that
in the normal one. In other words, its general form (but, of course, not the
concrete values of the coefficients $F,U,V$) does not depend on the
conformal frame, as it should be.

\section{Comparison to normal extreme and non-extreme black holes}

It is instructive to compare the modified action with that for extreme black
holes in the normal case and with the non-extreme one.

Treatment of black holes topologies in \cite{teit} (that extended the
approach of \cite{zan} to include extremal ones) revealed that there are
terms on the horizon of nonextreme black holes which are absent in the
extreme case. On the first glance, this exhibits close analogy with the role
of an inner boundary on the horizon in our situation, because of which it
would be tempting to interpret our approach simply as reformulation of the
well-known results. Actually, however, there is a big difference here since
not only the {\it value} of the action, but also its {\it form} should be
modified from the outset in the ''anomalous'' case (we are speak about the
Hilbert action or its generalization to the dilaton case). This can be
explained as follows.

We can rewrite the action (\ref{itot}), (\ref{i}) for the normal case as 
\begin{equation}
I_{gd}=\frac{1}{16\pi }\int d^{3}x\sqrt{g_{3}}\beta bZ_{0}^{0}+B-S\text{,}
\label{t}
\end{equation}
where $B=\int_{B}d\sigma \beta \varepsilon $, 
\begin{equation}
S=\frac{A_{eff}}{8}\chi \text{,}  \label{eu}
\end{equation}
$\chi =\frac{2T_{H}}{T_{0}}$. The value of $\chi $ coincides with the Euler
characteristic of the manifold, $\chi =2$ for non-extreme black holes and $%
\chi =0$ for extreme ones. In the case $F=1$ eq. (\ref{t}) coincides with
eqs. (12), (18) of \cite{teit}, if one identifies $I_{can}$ with $\frac{1}{%
16\pi }\int d^{3}x\sqrt{g_{3}}\beta bZ_{0}^{0}$, $B_{\infty }$ with $B$ (if
a boundary moves to infinity), $I_{H}$ with $I_{gd}$. Thus, for the normal
case the conclusion $S=0$ for the extreme black holes is made on the basis
of the {\it standard }Euclidean Hilbert action (\ref{ac}), (\ref{b}).

However, in our situation the action is modified: 
\begin{equation}
\tilde{I}_{gd}=I_{gd}+I_{1}\text{, }I_{1}\equiv \frac{1}{8\pi }\int_{H}d^{3}x%
\sqrt{g_{3}}FK\text{.}  \label{imod}
\end{equation}
Taking into account that $\sqrt{g_{3}}=b\sqrt{g_{2}}$ and using again the
relation (\ref{k2k}), we see that for ''normal'' extreme black holes, when $%
b_{,i}n^{i}\rightarrow 0$, $I_{1}$ vanishes. However, it does not vanish in
the anomalous case and it is just this term that regularizes the action and
gives the well-defined value for the entropy.

For non-extreme topologies one may also choose to place an inner boundary
and, as a horizon is screened in this case for an external observer, he
would not detect any entropy. However, there is a big difference here
between such a system and our anomalous black hole. For the non-extreme case
one can put an inner boundary in any place between a horizon and an outer
boundary, an inner shell being physical in the sense it is built up from
matter. Then the action has the general form (\ref{b}) but its concrete
value changes due to adding this shell.

Meanwhile, for anomalous extreme black holes introducing an inner boundary
is mandatory. This boundary should be place on a horizon but the ''shell''
itself is fictitious. This is simply the way to express the fact that now
the action has the modified form (\ref{imod}), which take into account the
contribution from the horizon.

To summarize, there are three typical situations: (1) non-extreme black
holes, (2) ''normal'' extreme black holes, (3) ''anomalous'' extreme black
holes. The {\it form} of the Hilbert action in the cases (1) and (2) is the
same, but the {\it value }of the entropy is different $(S=A_{eff}/4$ and $%
S=0 $); in the case (3) the {\it form} of the Hilbert action (more exactly,
its generalization to the dilaton case) differs from that in (1), (2) but
the {\it value} of the entropy is the same as in (2). The papers \cite{ross}%
, \cite{teit}, \cite{gib} make accent on difference between (1) and (2),
whereas the present article - on difference between (2)\ and (3).

\section{Exact solutions}

In this section we consider, as examples, some exact solutions of
self-consistent scalar-gravity theories. In all case the metric is
spherically-symmetrical: 
\begin{equation}
ds^{2}=d\tau ^{2}b^{2}+\alpha ^{2}dy^{2}+r^{2}(y)(d\theta ^{2}+\sin
^{2}\theta d\phi ^{2})\text{.}  \label{y}
\end{equation}
It follows from the standard relations for the Hawking temperature $T_{H}=%
\frac{\kappa }{2\pi }$, $\kappa =\lim \frac{\partial b}{\partial l}$
(hererafter the sign ''$\lim $'' refers to the horizon) that 
\begin{equation}
T_{H}=\frac{1}{2\pi }\lim \alpha ^{-1}\frac{\partial b}{\partial y}.
\label{thaw}
\end{equation}

The formula for the energy takes a very simple form: $E=4\pi \varepsilon
r^{2}=A_{eff}\varepsilon _{eff}$, where $A_{eff}=4\pi r^{2}F$ and, according
to (\ref{e}), (\ref{e2}) 
\begin{equation}
\varepsilon _{eff}=-\frac{1}{8\pi }\frac{(Fr^{2})^{\prime }}{Fr^{2}\alpha }%
\text{,}
\end{equation}
\begin{equation}
\varepsilon =\varepsilon _{eff}F\text{,}
\end{equation}
\begin{equation}
E=-\frac{1}{2}\frac{(Fr^{2})^{\prime }}{\alpha }\text{,}  \label{en}
\end{equation}
prime denotes differentiation with respect to $y$. In a similar way, 
\begin{equation}
Y_{1}=\lim \frac{\beta _{0}br^{2}F^{\prime }}{2\alpha }\text{,}  \label{1}
\end{equation}
\begin{equation}
Y_{2}=\lim \frac{\beta _{0}b^{\prime }r^{2}F}{2\alpha }\text{,}  \label{2}
\end{equation}
\begin{equation}
\Delta I=-\frac{3}{2}\beta _{0}\psi ^{\prime }(\phi )Y_{3}\text{, }%
Y_{3}=\lim \frac{r^{2}bF\phi ^{\prime }}{\alpha }\text{.}  \label{3}
\end{equation}
It is assumed that the $r$ coordinate runs from smaller to larger values
from a horizon towards infinity.

The modified gravitation-dilaton Euclidean action is 
\begin{equation}
\tilde{I}_{gd}=\beta _{B}E_{B}-\beta _{H}^{loc}E_{H}\text{,}  \label{atot}
\end{equation}
$\beta ^{loc}\,$is the inverse local temperature that tends to zero on the
horizon$.$ We will see below that, as one approaches the horizon, $E_{H}$
tends to infinity but, nevertheless, in some cases the product $\beta
_{H}^{loc}E_{H}$ remains finite.

\subsection{Black holes in Brans-Dicke theory}

Consider the Brans-Dicke theory, for which $F=\phi $, $V=-\omega \phi ^{-1}$%
, $U=0$ ($\omega =const$). There exist exact solutions within this theory,
describing black holes \cite{jor} - \cite{cold}. They fall into two classes.
Consider the first one, using notations of (\ref{y}):

\subsubsection{Case 1}

\begin{equation}
b=z^{(Q-\chi )/2}\text{, }\alpha =z^{-Q/2}\text{, }r^{2}=y^{2}z^{1-Q}\text{, 
}z=(1-\frac{y_{+}}{y})\text{.}  \label{11}
\end{equation}
\begin{equation}
\phi =z^{\chi /2}\text{.}  \label{12}
\end{equation}
It is supposed that 
\begin{equation}
\text{ }Q>\chi \text{, }Q\geq 2\text{.}  \label{q}
\end{equation}
The first condition in (\ref{q}) ensures that $y=y_{+}$ is a horizon, the
second follows from its regularity (finiteness of the Kretschmann invariant (%
\ref{inv}), see \cite{cold0} for details).

We get from (\ref{thaw}) that 
\begin{equation}
T_{H}=\frac{Q-\chi }{8\pi y_{+}}\lim_{z\rightarrow 0}z^{\gamma /2}\text{, }%
\gamma \equiv 2Q-\chi -2\text{.}  \label{th1}
\end{equation}
It follows directly from (\ref{q}) that $\gamma >0$. Thus, $T_{H}=0$ as a
direct consequence of the regularity conditions.

Direct calculations give us that $Y_{3}\sim z^{1+\chi /2}$on the horizon, so 
$Y_{3}$ diverges for $\chi <-2$. We also get 
\begin{equation}
Y_{1}=\frac{\beta _{0}y_{+}\chi }{4}\text{, }Y_{2}=\frac{\beta _{0}y_{+}}{4}%
(Q-\chi )\text{.}  \label{1y12}
\end{equation}
Apart from this, on the horizon 
\begin{equation}
A_{eff}\simeq 4\pi y_{+}^{2}z^{1-Q+\chi /2}\rightarrow \infty \text{,}
\end{equation}
where we took into account the inequality $Q>1+\chi /2$ that follows from (%
\ref{q}). The energy density 
\begin{equation}
\varepsilon _{eff}=\frac{Q-1-\chi /2}{8\pi y_{+}}z^{Q/2-1}\text{,}
\end{equation}
$\varepsilon \sim z^{(Q+\chi )/2-1}$. Thus, on the horizon $z=0$ $%
\varepsilon _{eff}$ remains finite or even vanish due to the regularity
condition (\ref{q}), while $\varepsilon $ may diverge for negative $\chi $
large enough. It follows from eq. (\ref{en}) that 
\begin{equation}
\beta _{H}^{loc}E_{H}=\frac{\beta _{0}y_{+}}{2}(Q-\frac{\chi }{2}-1)\text{.}
\end{equation}

\subsubsection{Comparison with the results of \protect\cite{kim}}

Thermodynamics of black holes described by the exact solutions (\ref{11}), (%
\ref{12}) was discussed in \cite{kim}. In the space of parameters the
authors considered three cases. Not counting the trivial case of the the
Schwarzschild metric ($Q=1$, $\chi =0$, $\phi =const$), they considered two
cases for which they obtained $T_{H}=\infty ,$ whereas we get $T_{H}=0$. As
this value for the Hawking temperature contrasts sharply with what is
obtained in our article, we will dwell upon on the reason of this
discrepancy. The paramters of the metric (\ref{11}) obey the condition
(eq.(4) of \cite{kim}) 
\begin{equation}
Q^{2}+(1+\frac{\omega }{2})\chi ^{2}-Q\chi -1=0\text{.}  \label{qw}
\end{equation}
Here $\omega $ is the Brans-Dicke parameter. Both cases I and II considered
in \cite{kim} correspond to $\omega +\frac{3}{2}>0$. It follows directly
from (\ref{qw}) that in this case the parameter $\gamma $ that appears in (%
\ref{th1}) is negative and one gets formally $T_{H}=\infty $. However, this
contradicts the regularity condition (\ref{q}). This also conflicts with the
incorrect statement made in \cite{kim} (in discussion after listing formulas
with $T_{H}=\infty $) that black holes with $Q-\frac{\chi }{2}<1$ are
regular. (But the authors themselves point out rightly that in their case II
the horizon is singular.) In our view, thermodynamics of black holes with a
singular horizon has no physical meaning at all. For example, if the
finiteness of the invariant (\ref{inv}) is relaxed, proof of the constancy
of the surface gravity on the horizon (the condition $\rho _{,a}=0$) loses
its sense, the zero law of thermodynamics fails and one cannot even
introduce the notion of a black hole temperature. Therefore, thermodynamic
approach to such objects does not apply and not only the claim made in \cite
{kim} about the value $S=0$ but also the notion of the entropy itself is no
longer valid under these circumstances. In other words, ''thermodynamics''
is considered in \cite{kim} in the range of parameters, where there are no
thermodynamic objects.

On the other hand, in the complimentary range of parameters ($\omega +\frac{3%
}{2}<0$ or, equivalently, (\ref{q})), omitted in \cite{kim}, thermodynamic
properties of black holes are well defined, provided the Euclidean approach
is properly modified.

\subsubsection{Case 2}

\begin{equation}
b=\exp [-(c+\frac{s}{2})y]\text{, }\alpha =y^{-2}\exp [(c-\frac{s}{2})y]%
\text{, }r^{2}=y^{-2}\exp [(2c-s)y]\text{.}
\end{equation}
\begin{equation}
\phi =e^{sy}\text{.}
\end{equation}

Here $2c-s>0$, $c>0$, $-2c<s<2c$ (see \cite{cold0} for details). The horizon
lies at $y\rightarrow \infty $. The formula (\ref{thaw}) gives us 
\begin{equation}
T_{H}=\frac{s+2c}{8\pi }\lim_{y\rightarrow \infty }y^{2}\exp (-2cy)=0\text{.}
\end{equation}

Again, direct calculations shows that $Y_{1}=-\frac{\beta _{0}s}{2}$, $Y_{2}=%
\frac{s+2c}{2s}$, $Y_{3}$ diverges as $e^{sy}$, 
\begin{equation}
A_{eff}=4\pi y^{-2}\exp (2cy)\rightarrow \infty \text{,}
\end{equation}
on the horizon 
\begin{equation}
\varepsilon _{eff}=\frac{cy^{2}}{4\pi }\exp [\left( \frac{s-2c}{2}\right)
y]\rightarrow 0\text{,}
\end{equation}
\begin{equation}
\varepsilon \sim y^{2}\exp [\left( \frac{3s-2c}{2}\right) y]
\end{equation}
may be finite or infinite dependent on the relation between $s$ and $c$, 
\begin{equation}
\beta _{H}^{loc}E_{H}=\beta _{0}c\text{.}
\end{equation}

\subsection{BBMB solution}

This is exact solution for the coupling $F=1-\xi \phi ^{2}$, where $\xi $
corresponds to the conformal case \cite{bbm}. It represents the rare
exception, when a black hole with a finite area remains regular in spite of
the presence of scalar hair. Its form coincides with the extremal
Reissner-Nordstr\"{o}m one:

\begin{equation}
b=(1-\frac{M}{r})\text{, }\alpha =b^{-1}\text{, }r(y)=y\text{, }T_{H}=0\text{%
,}
\end{equation}
\begin{equation}
\phi =q(r-M)^{-1}\text{, }M=\left( \frac{4\pi q^{2}}{3}\right) ^{1/2}\text{.}
\end{equation}
In so doing, this hair is a discrete one, manifesting itself in a choice of
the sign of $\phi $ since for a given mass $M$ there are two possible values
of $q.$ It was shown in \cite{bek75} that, in spite of divergencies in $F$
and $\phi $ on the horizon, nothing pathological occurs with a particle,
approaching the horizon, even if it is coupled to $\phi $. However, whereas
mechanics remains well-defined, the standard approach would fail for
thermodynamics since in the present case $Y_{1}=\beta _{0}\xi
q^{2}(r-M)^{-1} $, $Y_{2}=-\frac{1}{2}\beta _{0}\xi q^{2}(r-M)^{-1}$, $Y_{3}$
diverges as $(r-M)^{-2}$, $A_{eff}=-4\pi M^{2}\xi q^{2}(r-M)^{-2}$ $%
\rightarrow -\infty $.

The attempt to apply the standard formalism here leads to the meaningless
result - entropy that not only diverges, but even is negative since $Y_{2}<0$%
. This drawback can be repaired according to the prescription, described
above with the result $S=0$. However, a new difficulty arises here. Simple
calculations show that near the horizon 
\begin{equation}
\varepsilon _{eff}=(4\pi M)^{-1}
\end{equation}
and 
\begin{equation}
\beta _{H}^{loc}E_{H}=-\beta _{0}q^{2}\xi (r-M)^{-1}\text{.}
\end{equation}
In so doing, the total Euclidean action, according to (\ref{atot}), $I\sim
(r-M)^{-1}\rightarrow +\infty $.

\subsection{Common and distinct features of all three solutions}

It is convenient to summarize relevant characteristics of our solutions in
one table:

\begin{tabular}{|l|l|l|l|}
\hline
& Brans-Dicke (case 1) & Brans-Dicke (case 2) & BBMB \\ \hline
$Y_{1}$ & $\frac{\beta _{0}y_{+}\chi }{4}$ & $-\frac{\beta _{0}s}{2}$ & $%
+\infty $ \\ \hline
$Y_{2}$ & $\frac{\beta _{0}y_{+}}{4}(Q-\chi )>0$ & $\frac{s+2c}{2s}>0$ & $%
-\infty $ \\ \hline
$Y_{3}$ & infinite or finite depending on parameters & infinite & infinite
\\ \hline
$A_{eff}$ & $+\infty $ & $+\infty $ & $-\infty $ \\ \hline
$T_{H}$ & $0$ & $0$ & $0$ \\ \hline
$S$ & $0$ & $0$ & $0$ \\ \hline
$I$ & finite & finite & $+\infty $ \\ \hline
$\tilde{I}$ & finite & finite & $+\infty $ \\ \hline
\end{tabular}

Here we did not display the sign of $Y_{3}$ since it is unimportant in the
given context.

That the standard approach gives an unsatisfactory answer for the BBMB
solution follows from divergencies in the action. It is also obvious that $%
Y_{2}$ cannot be considered as a candidat on the BBMB entropy because $%
Y_{2}\rightarrow -\infty $. For Brans-Dicke black holes the action is finite
but the entropy, if identified with $Y_{2}$ according to (\ref{sef}), would
also have given an unphysical result. Indeed, $Y_{2}$ is model dependent and
would be imcompatible with the relation between the Euler characteristics
and entropy \cite{teit}, \cite{gib}, \cite{zan} in contrast to the universal
form $S=0$ for ''normal'' extreme black holes.

The fact that $Y_{1}\neq 0$, explains the roots of these difficulties.
Indeed, let us, according to the general formula (\ref{y1}), write down $%
Y_{1}$ in the spherically-symmetrical case, singling out the factor $b$, as $%
Y_{1}=y_{1}b$. Then the variation with respect to $b$ gives us $\delta
Y_{1}=y_{1}\delta b$. If $Y_{1}$ is finite, near the horizon $y_{1}\sim
b^{-1}$. In the variation procedure the function $b$ and its variation $%
\delta b$ within the same class behave in a similar way, as one approaches
the horizon. As a result, $\delta Y_{1}\neq 0$ and, according to (\ref{di}),
the variational procedure fails and one cannot state that the Euclidean
metrics under consideration were obtained in the self-consistent way (in the
normal case we would have finite $y_{1}$, so the product $y_{1}\delta
b\rightarrow 0$ due to the behavior of $b$ near the horizon). In the case of
BBMB solutions the situation even gets worse since $Y_{1}$ diverges. Thus,
the standard Euclidean action formalism fails in all three cases.

As far as the modification (treating a horizon as an inner noundary with the
corresponding terms in the action) is concerned, we see deep distinction
between Brans-Dicke black holes and the BBMB\ solution. For both types of
Brans-Dicke black holes black holes (i) the black hole entropy $S$ is
well-defined, $S=0$; (ii) the energy associated with a horizon is infinite
but the Euclidean action itself is finite. The latter reveals itself in the
finite nonzero product $\beta _{H}^{loc}E_{H}$, where the first term tends
to zero, while the second one diverges.

However, for the BBMB solution, in spite of the fact that we found formally $%
S=0$, the total action turns out to be infinite and positive both with or
without a boundary on the horizon. This means that such solution cannot
contribute into the partition function $Z\backsim \exp (-I)$, so
thermodynamics (if any), which can be assigned to such solutions, is very
poor. It is seen directly from the table that some properties of the BBMB
black holes are in a sense ''more peculiar'' than those of the Brans-Dicke
ones. Indeed, both $A_{eff}$ and $Y_{2}$ (analogues of the quantities,
proportional to the entropy of ''normal'' black holes) are negative, and
these oddities make it impossible to restore reasonable thermodynamic
properties even after the ''improvement'' of the action. The failure of
thermodynamic interpretation for the BBMB black holes can be ascribed to the
fact that, with a finite qualilocal energy density on the horizon $%
\varepsilon _{eff}$, the effective area $A_{eff}$ as well as energy $%
\varepsilon _{eff}A_{eff}$ grows more rapidly than the local inverse Tolman
temperature tends to zero.

\section{Summary and conclusion}

It is shown, without using the assumption of the spherical symmetry, that
any static black hole with an infinite horizon area but regular horizon
should have a zero Hawking temperature. It follows from our consideration
that black holes with an infinite effective horizon area occupy, formally
speaking, an intermediate place between non-extreme and ''normal'' extreme
black holes in what concerns thermodynamics. Formally, it is seen from (\ref
{sef}): the factor $\frac{T_{H}}{T_{0}}\rightarrow 0$ (what is typical of
extreme black holes), but the factor $\frac{A_{eff}}{4}$ (typical of
non-extreme ones) remains important and even tends to infinity. As a result,
their product $Y_{2}$ may be finite (as is the case for Brans-Dicke black
holes discussed above). However, the true situation is even more complicated
than this rough analogy since, as we saw, one cannot identify (\ref{sef})
with the entropy. Moreover, the corresponding quantity can be negative and
diverge as it happens to BBMB solutions.

We traced some subtleties in the action principle for such unusual
geometries and demonstrated that the action formalism, provided it is
modified properly, handles even rather exotic situations, when either the
horizon area or the dilaton coefficient $F$ diverges. However, this does not
guarantee in advance that black holes in {\it any} theory of this kind
represent well-defined thermodynamic objects. Explicit examination of exact
solutions showed that this is the case for Brance-Dicke black holes but not
for the BBMB\ solution.

As far as Brans-Dicke theory is conserned, is just the unusual character of
black holes under consideration (infinite $A_{eff}$) that makes the link
between conformal properties of the action, self-consistency of the
variational procedure and the zero value of the black hole entropy
nontrivial. The Euclidean action is finite in spite of the fact that the
quasilocal energy $E_{H}$ associated with a horizon diverges. These
divergencies receive a simple explanation: they appear due to an infinite
effective area $A_{eff}$, while the effective energy $\varepsilon _{eff}$
per unit effective area turns out to be finite or even vanish.

The key point in our treatment consisted in placing an additional inner
boundary on the horizon of extremal black holes with infinite $A_{eff}$. The
reason, why this was necessary, consists in failure of the variational
procedure without corresponding boundary terms. This failure would reveal
itself in the appearance of superfluous terms which, in particular, would
contain the normal derivatives of the local temperature. They are
automatically equal to zero in a normal case (finite $A_{eff}$) due to
properties of the horizon but, in general, persist in our case (infinite $%
A_{eff}$). The suggested approach removes all these undesirable terms and
gives a quite definite answer for the black hole entropy.

On the other hand, failure of thermodynamic interpertation of the BBMB black
holes is, in our view, especially interesting. We have already paid
attention that thermodynamics interpretation of some models of
two-dimensional dilaton gravity fails \cite{nonext}, \cite{reg2d}. This
turns out to be possible due to quantum effects entirely and refer to
nonextreme horizons. Now we see that, in an essense, BBMB black hole can be
thought of as a classical extremal counterpart of such exceptional solutions.

On the other hand, it is also instructive to carry out some parallels
bewteen semiclassical two-dimensional dilaton theories and pure classical
Branse-Dicke black holes considered in the present paper. As is shown in 
\cite{nonext}, \cite{reg2d}, \cite{ext}, when the quantum-corrected quantity 
$F$ diverges on the horizon, in some special two-dimensional models infinite
backreaction remains compatible with regularity of the geometry. As the
quantity $F$ of two-dimensional models, obtained by sperically-symmetrical
reduction, is similar to $r^{2}$, divergencies in $F$ resemble an infinite
horzion area of four-dimensional ones. Thus, although the results of the
present paper are restricted to the classical domain only, the
aforementioned analogy seems to testify that at least for some models the
value of the black hole entropy $S=0$, inherent to classical extremal black
holes, survives with quantum backreaction taken into account.

When $A_{eff}$ is inifnite, the issue of quantum backreaction in so unusual
situation deserves separate treatment. Anyway, it was necessary, as the
first step, to elaborate self-closed and self-consistent approach to
thermodynamics of objects with infinite $A_{eff}$ within pure classical
framework, and this task is performed in the present article.





%
%

%
%

\end{document}